\documentclass[pre,aps,preprint,showpacs,preprintnumbers,amsmath,amssymb,secnumarabic,eqsecnum]{revtex4}

\usepackage{graphicx}
\usepackage{dcolumn}
\usepackage{bm}

\newcommand{\beq}{\begin{equation}}
\newcommand{\eeq}{\end{equation}}
\newcommand{\beqa}{\begin{eqnarray}}
\newcommand{\eeqa}{\end{eqnarray}}
\newcommand{\vc}[1]{\mbox{\boldmath $#1$}}

\newcommand{\vol}[1]{{\bf #1}}

\newcommand{\du}[1]{{\bf\sf #1}}

\begin{document}


\title{Efficient swimming of an assembly of rigid spheres at low Reynolds number}

\author{B. U. Felderhof}

 \email{ufelder@physik.rwth-aachen.de}
\affiliation{Institut f\"ur Theorie der Statistischen Physik \\ RWTH Aachen University\\
Templergraben 55\\52056 Aachen\\ Germany\\
}%



\date{\today}

\begin{abstract}
The swimming of an assembly of rigid spheres immersed in a viscous fluid of infinite extent is studied in low Reynolds number hydrodynamics. The instantaneous swimming velocity and rate of dissipation are expressed in terms of the time-dependent displacements of sphere centers about their collective motion. For small amplitude swimming with periodically oscillating displacements, optimization of the mean swimming speed at given mean power leads to an eigenvalue problem involving a velocity matrix and a power matrix. The corresponding optimal stroke permits generalization to large amplitude motion in a model of spheres with harmonic interactions and corresponding actuating forces. The method allows straightforward calculation of the swimming performance of structures modeled as assemblies of interacting rigid spheres. A model of three collinear spheres with motion along the common axis is studied as an example.
\end{abstract}

\pacs{47.15.G-, 47.63.mf, 47.63.Gd, 45.50.Jf}
\maketitle
\section{\label{1}Introduction}

In earlier work \cite{1} we presented a method to analyze the performance of a microswimmer modeled as an assembly of $N$ rigid spheres immersed in a viscous incompressible fluid of infinite extent, with a no-slip boundary condition on the surface of each sphere. The motion of the whole system is determined by the Stokes equations of low Reynolds number hydrodynamics. The swimming motion of such a system was discussed earlier by Alouges et al. \cite{2},\cite{3}. The particular case of collinear spheres was studied by Vladimirov \cite{4} using a two-timing method.

For small displacements of the spheres from fixed positions in the collective rest frame the time-averaged swimming velocity and rate of dissipation can be evaluated in terms of a $(3N-3)\times(3N-3)$ velocity matrix and a $(3N-3)\times(3N-3)$ power matrix, which can be constructed from the mobility matrix for each relative rest configuration \cite{1}. Optimization of the velocity at fixed power leads to a generalized eigenvalue problem involving the two matrices. Optimal efficiency corresponds to the maximum eigenvalue.

In a model with harmonically interacting spheres the optimal stroke of small amplitude motion can be used to calculate a set of corresponding actuating forces. Large amplitude motion can be studied by solving the equations of Stokesian dynamics for the same actuating forces multiplied by a factor. The mean swimming velocity and the mean power of the large amplitude motion can then be determined numerically from the limit cycle of the solution.

In the following we present an alternative method based on a purely kinematic point of view. Expressions are derived for the instantaneous swimming velocity and power in terms of the sphere displacements from the center and their instantaneous time derivative. This allows calculation of the mean swimming velocity and mean power for given periodic stroke of any amplitude. The present method also provides an alternative derivation of the velocity matrix and power matrix of small amplitude motion.

For large amplitude swimming the present method is more straightforward than the earlier one \cite{1}, since it does not require numerical solution of the equations of Stokesian dynamics. A large amplitude stroke may be determined by amplifying the optimal stroke found from the eigenvalue problem of the small amplitude theory for a given equilibrium structure. The instantaneous swimming velocity and power are then determined from explicit expressions in terms of the given displacements. Subsequently the mean swimming velocity and mean power can be found by integration over a period.

Both methods are tested on a model of three collinear spheres with motion along the common axis, as formulated by Najafi and Golestanian \cite{5} and studied in detail by Golestanian and Ajdari \cite{6}. The two methods of calculation lead to similar numerical results for a wide range of amplitude.

\section{\label{2}Displacement and swimming velocity}

We consider a set of $N$ rigid spheres of radii $a_1,...,a_N$ immersed in a viscous
incompressible fluid of shear viscosity $\eta$. The fluid is of infinite extent in all directions. At low Reynolds
number and on a slow time scale the flow velocity
$\vc{v}$ and the pressure $p$ satisfy the
Stokes equations \cite{7}
\begin{equation}
\label{2.1}\eta\nabla^2\vc{v}-\nabla p=0,\qquad\nabla\cdot\vc{v}=0.
\end{equation}
The flow velocity $\vc{v}$ is assumed to satisfy the no-slip boundary condition on the surface of the spheres.
The fluid is set in motion by time-dependent motions of the
spheres. At each time $t$ the velocity field $\vc{v}(\vc{r},t)$ tends to zero at infinity, and the pressure $p(\vc{r},t)$ tends to the constant ambient pressure $p_0$.
We shall study periodic relative motions which lead to swimming
motion of the collection of spheres.

We assume that the motion is caused by time-dependent periodic forces $\vc{F}_1(t),...,\vc{F}_N(t)$ which satisfy the condition that their sum vanishes at any time. The forces are transmitted by the spheres to the fluid. The spheres can rotate freely, so that they exert no torques on the fluid. Hence the rotational velocities $\vc{\Omega}_1(t),..., \vc{\Omega}_N(t)$ can be ignored. The translational velocities $\vc{U}_1,...,\vc{U}_N$ are linearly related to the forces,
\begin{equation}
\label{2.2}\vc{U}_j=\sum^N_{k=1}\vc{\mu}^{tt}_{jk}\cdot\vc{F}_k,\qquad j=1,...,N,
\end{equation}
 with translational mobility tensors $\vc{\mu}^{tt}_{jk}$. The tensors have many-body character and depend in principle on the positions of all particles  \cite{8}-\cite{10}. By translational invariance only relative distance vectors $\{\vc{R}_i-\vc{R}_j\}$ occur in the functional dependence. We abbreviate eq. (2.2) as
 \begin{equation}
\label{2.3}\du{U}=\vc{\mu}\cdot\du{F},
\end{equation}
with a symmetric $3N\times 3N$ mobility matrix $\vc{\mu}$. Conversely
 \begin{equation}
\label{2.4}\du{F}=\vc{\zeta}\cdot\du{U},
\end{equation}
with friction matrix $\vc{\zeta}$. The friction matrix is the inverse of the mobility matrix, $\vc{\zeta}=\vc{\mu}^{-1}$,
and is also symmetric.

The positions of the centers change as a function of time. The equations of motion of Stokesian dynamics read
  \begin{equation}
\label{2.5}\frac{d\vc{R}_j}{dt}=\vc{U}_j(\vc{R}_1,...,\vc{R}_N,t),\qquad j=1,...,N.
\end{equation}
The explicit time-dependence on the right originates in the time-dependence of the forces $\du{F}(t)$. In the swimming motion the forces are periodic in time with period $T$, so that $\du{F}(t+T)=\du{F}(t)$. As mentioned, we impose the condition that at no time is there a net force acting on the set of spheres, so that
  \begin{equation}
\label{2.6}\sum^{N}_{j=1}\vc{F}_j(t)=0.
\end{equation}
We look for a solution of eq. (2.5) corresponding to swimming motion, of the form
\begin{equation}
\label{2.7}\vc{R}_j(t)=\vc{S}_{j0}+\int^t_0\vc{U}(t')\;dt'+\vc{\delta}_j(t),\qquad j=1,...,N,
\end{equation}
where the first two terms describe the collective motion of the configuration $\du{S}_0=(\vc{S}_{10},...,\vc{S}_{N0})$ with swimming velocity $\vc{U}(t)$ caused by the displacements $\{\vc{\delta}_j(t)\}$. We require that the latter are periodic with period $T$, and exclude uniform displacements, so that the $3N$-dimensional vector $\du{d}(t)=\{ \vc{\delta}_1(t),...,\vc{\delta}_N(t)\}$ satisfies
\begin{equation}
\label{2.8}\du{d}(t)\cdot\du{u}_\alpha=0,\qquad(\alpha=x,y,z),
\end{equation}
where the symbol $\du{u}_x$ denotes a $3N$-dimensional vector with $1$ on the $x$ positions, $0$ on the $y,z$ positions, and cyclic. Periodicity implies
\begin{equation}
\label{2.9}\vc{U}(t+T)=\vc{U}(t),\qquad\du{d}(t+T)=\du{d}(t).
\end{equation}
The mean swimming velocity is defined as
\begin{equation}
\label{2.10}\overline{\vc{U}}_{sw}=\frac{1}{T}\int^T_0\vc{U}(t)\;dt.
\end{equation}
We require that $\du{d}(t)$ is purely oscillating, so that
\begin{equation}
\label{2.11}\int^T_0\du{d}(t)\;dt=\du{0}.
\end{equation}

We show in the following that the instantaneous swimming velocity $\vc{U}(t)$ can be calculated from the displacement vector $\du{d}(t)$ and its time derivative $\dot{\du{d}}(t)$. Later we compare the present kinematic description to a dynamical model, in which the forces are decomposed into actuating forces and elastic restoring forces.

\section{\label{3}Swimming velocity and dissipation}

By substitution of eq. (2.7) into eqs. (2.4) and (2.5) one finds
\begin{equation}
\label{3.1}\du{F}=\vc{\zeta}\cdot(U_\beta\du{u}_\beta+\dot{\du{d}}),
\end{equation}
where summation over repeated greek indices is implied. The condition (2.6) can be expressed as $\du{u}_\alpha\cdot\du{F}=0$, so that
\begin{equation}
\label{3.2}Z_{\alpha\beta}U_\beta=-\du{u}_\alpha\cdot\vc{\zeta}\cdot\dot{\du{d}}
\end{equation}
with friction tensor
\begin{equation}
\label{3.3}Z_{\alpha\beta}=\du{u}_\alpha\cdot\vc{\zeta}\cdot\du{u}_\beta.
\end{equation}
Hence we obtain the swimming velocity
\begin{equation}
\label{3.4}U_\alpha=-M_{\alpha\beta}\du{u}_\beta\cdot\vc{\zeta}\cdot\dot{\du{d}},
\end{equation}
where $M_{\alpha\beta}$ is the inverse of the friction tensor. The $3N\times 3N$ friction  matrix $\vc{\zeta}$ depends only on the instantaneous relative positions. Therefore the friction tensor $\vc{Z}$ and the mobility tensor $\vc{M}$ depend on the displacement vector $\du{d}$, but not on the central coordinates $R_\alpha=\du{u}_\alpha\cdot\du{R}/N$.

By series expansion of the mobility tensor $\vc{M}$ and the friction matrix $\vc{\zeta}$ in powers of the displacement vector $\du{d}$ we obtain a corresponding expansion of the swimming velocity
\begin{equation}
\label{3.5}\vc{U}=\vc{U}^{(1)}+\vc{U}^{(2)}+\vc{U}^{(3)}+...,
\end{equation}
with first order term
\begin{equation}
\label{3.6}U^{(1)}_\alpha=-M^0_{\alpha\beta}\du{u}_\beta\cdot\vc{\zeta}^0\cdot\dot{\du{d}},
\end{equation}
with mobility tensor $M^0_{\alpha\beta}$ and friction matrix $\vc{\zeta}^0$ calculated for the configuration $\du{S}_0$. By periodicity of $\du{d}(t)$ the time average of the first order swimming velocity vanishes, $\overline{\vc{U}^{(1)}}=\vc{0}$.

We introduce the friction vectors
\begin{equation}
\label{3.7}\du{f}_\alpha=\du{u}_\alpha\cdot\vc{\zeta}=\vc{\zeta}\cdot\du{u}_\alpha,
\end{equation}
where we have used the symmetry of the friction matrix $\vc{\zeta}$. The vectors are related to the friction tensor by
\begin{equation}
\label{3.8}\du{u}_\alpha\cdot\du{f}_\beta=\du{u}_\beta\cdot\du{f}_\alpha=Z_{\alpha\beta}.
\end{equation}
From the Taylor series expansion of eq. (3.4) we find that the second order instantaneous swimming velocity can be expressed as
\begin{equation}
\label{3.9}U^{(2)}_\alpha=-\du{d}\cdot \du{V}^\alpha\big{|}_0\cdot\dot{\du{d}},
\end{equation}
with matrix $\du{V}^\alpha$ given by
\begin{equation}
\label{3.10}\du{V}^\alpha=\vc{\nabla}\big[M_{\alpha\beta}\du{f}_\beta\big],
\end{equation}
where $\vc{\nabla}$ is the gradient vector in $3N$-dimensional configuration space. The notation $\big|_0$ in eq. (3.9) indicates that the matrix-function is to be evaluated at $\du{R}=\du{S}_0$.

The expression on the right of eq. (3.10) may be written as a sum of two terms,
\begin{equation}
\label{3.11}\du{V}^\alpha=(\vc{\nabla}M_{\alpha\beta})\du{f}_\beta+M_{\alpha\beta}\du{D}^\beta,
\end{equation}
with derivative friction matrix
\begin{equation}
\label{3.12}\du{D}^\beta=\vc{\nabla}\du{f}_\beta.
\end{equation}
We introduce the gradient vectors
\begin{equation}
\label{3.13}\du{g}^\beta_\gamma=\du{D}^\beta\cdot\du{u}_\gamma=\vc{\nabla}Z_{\beta\gamma},
\end{equation}
and use the identity
\begin{equation}
\label{3.14}Z_{\alpha\gamma}M_{\gamma\beta}=\delta_{\alpha\beta}
\end{equation}
to show that
\begin{equation}
\label{3.15}\vc{\nabla}M_{\alpha\beta}=-M_{\alpha\gamma}\du{g}^\gamma_\delta M_{\delta\beta}.
\end{equation}
Then eq. (3.11) may be expressed alternatively as
\begin{equation}
\label{3.16}\du{V}^\alpha=M_{\alpha\beta}\breve{\du{D}}^{\beta},
\end{equation}
with reduced derivative friction matrix
\begin{equation}
\label{3.17}\breve{\du{D}}^{\beta}=\du{D}^\beta-\du{g}^\beta_\gamma M_{\gamma\delta}\du{f}_\delta.
\end{equation}
This matrix has the property
\begin{equation}
\label{3.18}\breve{\du{D}}^{\beta}\cdot\du{u}_\alpha=0.
\end{equation}
From the fact that $\vc{\zeta}$ depends only on relative coordinates it follows that $\du{u}_\alpha\cdot\vc{\nabla}\vc{\zeta}=\du{0}$, and hence
\begin{equation}
\label{3.19}\du{u}_\alpha\cdot\du{D}^{\beta}=\du{0},\qquad\du{u}_\alpha\cdot\du{g}^\beta_\gamma=0.
\end{equation}
As a consequence
\begin{equation}
\label{3.20}\du{u}_\alpha\cdot\du{V}^{\beta}=\du{0},\qquad\du{V}^\alpha\cdot\du{u}_\beta=\du{0}.
\end{equation}

 The time-dependent rate of dissipation can be expressed in the same matrix formalism. The rate of dissipation is given by
\begin{equation}
\label{3.21}\mathcal{D}=\du{F}\cdot\du{U}=\du{F}\cdot\dot{\du{d}},
\end{equation}
since $\du{F}\cdot\du{u}_\alpha=0$ on account of the condition eq. (2.6). Substituting eq. (3.1) we find
\begin{equation}
\label{3.22}\mathcal{D}=\dot{\du{d}}\cdot\vc{\zeta}\cdot\dot{\du{d}}+U_\alpha\dot{\du{d}}\cdot\du{f}_\alpha.
\end{equation}
It follows from eq. (3.4) that the rate of dissipation is at least of second order in $\du{d}$ and $\dot{\du{d}}$. To second order, by use of eq. (3.6),
\begin{equation}
\label{3.23}\mathcal{D}^{(2)}=\dot{\du{d}}\cdot\du{P}\cdot\dot{\du{d}}
\end{equation}
with matrix
\begin{equation}
\label{3.24}\du{P}=\vc{\zeta}^0-M^0_{\alpha\beta}\du{f}^0_\alpha\du{f}^0_\beta.
\end{equation}
The matrix is symmetric and has the properties
\begin{equation}
\label{3.25}\du{u}_\alpha\cdot\du{P}=\du{0},\qquad\du{P}\cdot\du{u}_\alpha=\du{0}.
\end{equation}
The properties eq. (3.20) and (3.25) allow us to reduce the dimension of the matrix description by three by the introduction of center and relative coordinates.

\section{\label{4}Velocity matrix vector and power matrix}
The center of the assembly is given by
\begin{equation}
\label{4.1}\vc{R}=\frac{1}{N}\sum_{j=1}^N\vc{R}_j=\frac{1}{N}\;\vc{e}_\alpha\du{u}_\alpha\cdot\du{R}
\end{equation}
with Cartesian unit vectors $\vc{e}_\alpha$. We define relative coordinates $\{\vc{r}_j\}$ as
  \begin{eqnarray}
\label{4.2}\vc{r}_1&=&\vc{R}_2-\vc{R}_1,\qquad\vc{r}_2=\vc{R}_3-\vc{R}_2,\qquad ...,\nonumber\\
\vc{r}_{N-1}&=&\vc{R}_N-\vc{R}_{N-1}, \qquad j=1,...,N-1,
\end{eqnarray}
and the corresponding $(3N-3)$-vector $\du{r}=(\vc{r}_1,...,\vc{r}_{N-1})$. The $3N$-vector $(\vc{R},\du{r})$ is related to the vector $\du{R}$ by a transformation matrix $\du{T}$ according to
\begin{equation}
\label{4.3}(\vc{R},\du{r})=\du{T}\cdot\du{R}
\end{equation}
with explicit form given by eqs. (4.1) and (4.2).

The matrices $\du{V}^\alpha$ and $\du{P}$ are transformed to
\begin{equation}
\label{4.4}\du{V}^\alpha_T=\du{T}\cdot\du{V}^\alpha\cdot\du{T}^{-1},\qquad\du{P}_T=\du{T}\cdot\du{P}\cdot\du{T}^{-1}.
\end{equation}
The first three rows of $\du{T}$ consist of $\du{u}_\alpha/N$ and the first three columns of $\du{T}^{-1}$ consist of $\du{u}_\alpha$. It follows from the properties eq. (3.20) and (3.25) that the first three rows and columns of the transformed matrices $\du{V}^\alpha_T$ and $\du{P}_T$ vanish identically. Hence in this representation we can drop the center coordinates and truncate the matrices by erasing the first three rows and columns. We denote the truncated $(3N-3)\times(3N-3)$-matrices as $\hat{\du{V}}_T^\alpha$ and $\hat{\du{P}}_T$ and define displacements $\vc{\xi}$ in relative space by
\begin{equation}
\label{4.5}(\vc{0},\vc{\xi})=\du{T}\cdot\du{d}.
\end{equation}
With this notation the second order swimming velocity and rate of dissipation are given by
\begin{equation}
\label{4.6}U^{(2)}_\alpha=\vc{\xi}\cdot\du{C}_T\cdot\hat{\du{V}}_T^\alpha\cdot\dot{\vc{\xi}},\qquad \mathcal{D}^{(2)}=\dot{\vc{\xi}}\cdot\du{C}_T\cdot\hat{\du{P}}_T\cdot\dot{\vc{\xi}},
\end{equation}
with the matrix
\begin{equation}
\label{4.7}\du{C}_T=[\widetilde{\du{T}^{-1}}\cdot\du{T}^{-1}]\;\vc{\hat{}}.
\end{equation}
This $(3N-3)\times(3N-3)$ dimensional matrix consists of numerical coefficients and is obtained from the corresponding $3N\times 3N$ matrix by truncation, as indicated by the final hat symbol.

We consider in particular harmonically varying displacements of the form
 \begin{equation}
\label{4.8}\du{d}(t)=\du{d}_s\sin\omega t+\du{d}_c\cos\omega t,
\end{equation}
 with a corresponding expression for $\vc{\xi}(t)$. The time-averaged second order swimming velocity and rate of dissipation are then given by
 \begin{eqnarray}
\label{4.9}\overline{U^{(2)}_\alpha}&=&\frac{1}{2}\omega\big[\vc{\xi}_s\cdot\du{C}_T\cdot\hat{\du{V}}_T^\alpha\big|_0\cdot\vc{\xi}_c
-\vc{\xi}_c\cdot\du{C}_T\cdot\hat{\du{V}}_T^\alpha\big|_0\cdot\vc{\xi}_s\big],\nonumber\\ \overline{\mathcal{D}^{(2)}}&=&\frac{1}{2}\omega^2\big[\vc{\xi}_s\cdot\du{C}_T\cdot\hat{\du{P}}_T\cdot\vc{\xi}_s+\vc{\xi}_c\cdot\du{C}_T\cdot\hat{\du{P}}_T\cdot\vc{\xi}_c\big].
\end{eqnarray}

We introduce the complex dimensionless vector
 \begin{equation}
\label{4.10}\vc{\xi}^c=\frac{1}{b}(\vc{\xi}_c+i\vc{\xi}_s),
 \end{equation}
 where $b$ is a typical length scale. With the definitions
  \begin{equation}
\label{4.11}\du{B}^\alpha=\frac{1}{2}ib\big(\du{C}_T\cdot\hat{\du{V}}_T^\alpha\big|_0-\widetilde{\du{C}_T\cdot\hat{\du{V}}_T^\alpha\big|_0}\big),\qquad
\du{A}=\frac{1}{b\eta}\;\du{C}_T\cdot\hat{\du{P}}_T,
\end{equation}
and the scalar product
  \begin{equation}
\label{4.12}(\vc{\xi}^c|\vc{\eta}^c)=\sum^{N-1}_{j=1}\vc{\xi}_j^{c*}\cdot\vc{\eta}^c_j
 \end{equation}
 the mean swimming velocity and mean rate of dissipation can then be expressed as
  \begin{equation}
\label{4.13}\overline{U^{(2)}_\alpha}=\frac{1}{2}\omega
b(\vc{\xi}^c|\du{B}^{\alpha}|\vc{\xi}^c),\qquad\overline{\mathcal{D}^{(2)}}=\frac{1}{2}\eta\omega^2b^3(\vc{\xi}^c|\du{A}|\vc{\xi}^c).
 \end{equation}
 We have normalized such that the matrix elements of $\du{B}^{\alpha}$ and $\du{A}$ are dimensionless. We call $\du{B}^{\alpha}$ the velocity matrix and $\du{A}$ the power matrix.

 We ask for the stroke with maximum swimming velocity in a class of strokes with equal rate of dissipation for fixed values of the geometric parameters, fixed frequency $\omega$, and fixed viscosity $\eta$. This leads to the generalized eigenvalue problem
 \begin{equation}
\label{4.14}\du{B}^\alpha\vc{\xi}^c=\lambda^\alpha\du{A}\vc{\xi}^c.
 \end{equation}
 The eigenvalues $\{\lambda^\alpha\}$ are real. The maximum efficiency for motion in direction $\alpha$ is given by the maximum eigenvalue as
   \begin{equation}
\label{4.15}E^\alpha_{Tmax}=\lambda^\alpha_{max}.
 \end{equation}
 The set $\{E^x_{Tmax},E^y_{Tmax},E^z_{Tmax}\}$ depends on the choice of Cartesian coordinate system. Further optimization may be possible by a rotation of axes. In particular cases a natural choice of axes will suggest itself.

 In the formulation of the mobility matrix in Eq. (2.2) the nature of the forces $\{\vc{F}_j\}$ need not be specified. In an earlier calculation \cite{11} we considered microswimmers with internal harmonic interactions, driven by actuating forces.
In matrix form the forces may be expressed as
 \begin{equation}
\label{4.16}\du{F}=\du{E}+\du{H}\cdot(\du{R}-\du{S}_0),
 \end{equation}
 with a real symmetric matrix $\du{H}$ with the property $\du{H}\cdot\du{u}_\alpha=0$. The actuating forces $\{\vc{E}_j(t)\}$ are assumed to satisfy
 \begin{equation}
\label{4.17}\sum_{j=1}^N\vc{E}_j(t)=0.
 \end{equation}
They can be generated internally or externally.

\section{\label{5}Three-sphere swimmer}

 The simplest application of the theory is to a three-sphere swimmer with three spheres aligned on the $x$ axis, as studied by Golestanian and Ajdari \cite{6}. The spheres move along the $x$ axis, and the $y$ and $z$ coordinates can be ignored. There are only two relative coordinates $r_1=x_2-x_1$ and $r_2=x_3-x_2$, and the relevant parts of the matrices $\du{B}^x$ and $\du{A}$ are two-dimensional. The elements of the $3\times 3$ mobility matrix are approximated by use of the Oseen interaction as \cite{7}
     \begin{equation}
\label{5.1}\mu^{tt}_{jk}=\frac{1}{6\pi\eta}\bigg[\frac{1}{a_j}\delta_{jk}+\frac{3}{2|x_j-x_k|}(1-\delta_{jk})\bigg].
 \end{equation}
 In the bilinear theory we consider a point $\du{r}_0$ in $\du{r}$-space with coordinates $(d_1,d_2)$, corresponding to the configuration $\du{S}_0$ of the rest system.
 As an example we consider the case of equal-sized spheres with $a_1=a_2=a_3=a$ and equal distances between centers $d_1=d_2=d$. For this case the explicit expressions for the matrices $\du{B}^x$ and $\du{A}$ are identical to those derived earlier by a different method \cite{1}. Explicit expressions for the eigenvectors $\vc{\xi}_\pm$ and eigenvalues $\lambda_\pm$ of the two-dimensional eigenvalue problem $\du{B}^x\cdot\vc{\xi}=\lambda\du{A}\vc{\xi}$, as functions of the ratio $d/a$, were derived in ref. 1.

In the bilinear theory, corresponding to small $\varepsilon$, the orbit $(r_1(t),r_2(t))=(x_2(t)-x_1(t),x_3(t)-x_2(t))$ in relative space is given by $\vc{r}(t)=\vc{r}_0+\vc{\xi}_0(t)$ with $\vc{r}_0=(d,d)$ and
\begin{equation}
\label{5.2}\vc{\xi}_0(t)=\varepsilon a\;\mathrm{Re}\;\vc{\xi}_+\exp(-i\omega t),
\end{equation}
with amplitude factor $\varepsilon$ and eigenvector $\vc{\xi}_+=(1,\xi_+)$ corresponding to the largest eigenvalue. In fig. 1 of ref. 1 we have shown the elliptical orbit in relative space for $d=5a$ and $\varepsilon=0.1$. The corresponding displacement vector in configuration space is given by
\begin{equation}
\label{5.3}\du{d}_0(t)=\du{T}^{-1}\cdot\left(\begin{array}{c}0\\\vc{\xi}_0(t)\end{array}\right),
\qquad\du{T}=\left(\begin{array}{ccc}\frac{1}{3}&\frac{1}{3}&\frac{1}{3}\\-1&1&0\\0&-1&1
\end{array}\right).
\end{equation}
In fig. 1 we show the reduced mean swimming velocity $\overline{U}_{sw}/(\varepsilon^2\omega a)$ as a function of $\varepsilon$ for $d=5a$, as calculated from eq. (3.4). In fig. 2 we show the reduced mean rate of dissipation $\overline{\mathcal{D}}/(\varepsilon^2\eta\omega^2a^3)$, as calculated from eq. (3.22). In fig. 3 we show the efficiency $E_T=\eta\omega a^2\overline{U}_{sw}/\overline{\mathcal{D}}$ as a function of $\varepsilon$. The efficiency increases monotonically with the amplitude factor.

It is of interest to compare the above results with values obtained by the numerical solution of the Stokesian equations of motion eq. (2.5) with hydrodynamic interactions given by eq. (5.1) and prescribed oscillating actuating forces. We use harmonic interactions given by the $3\times 3$-matrix
  \begin{equation}
\label{5.4}\du{H}=k\left(\begin{array}{ccc}-1&1&0\\1&-2&1\\0&1&-1
\end{array}\right)
\end{equation}
with elastic constant $k$. This corresponds to nearest neighbor interactions of equal strength $k$ between the three spheres. The stiffness of the swimmer is characterized by the dimensionless number $\sigma$ defined by
  \begin{equation}
\label{5.5}\sigma=\frac{k}{\pi\eta a\omega}.
\end{equation}

In general, the first order forces $\du{F}^{(1)}_0(t)$ corresponding to the displacement vector $\du{d}_0(t)$ and the corresponding first order swimming velocity $\vc{U}^{(1)}_0(t)$, calculated from eq. (3.6), follow from eq. (3.1) as
\begin{equation}
\label{5.6}\du{F}^{(1)}_0=\vc{\zeta}^0\cdot(U^{(1)}_{0\beta}\du{u}_\beta+\dot{\du{d}}_0).
\end{equation}
In the present case only the $x$ components are relevant.
 The corresponding actuating forces $\du{E}_0(t)$ are found from eq. (4.16) as
 \begin{equation}
\label{5.7}\du{E}_0(t)=\du{F}^{(1)}_0(t)-\du{H}\cdot\du{d}_0(t).
\end{equation}
These have the property $\du{u}_\alpha\cdot\du{E}_0(t)=0$, so that the sum of actuating forces vanishes. We choose initial conditions for the $x$ coordinates
 \begin{equation}
\label{5.8}x_1(0)=0,\qquad x_2(0)=d+\varepsilon a,\qquad x_3(0)=2d+\varepsilon a+\varepsilon a\;\mathrm{Re}\;\xi_+.
\end{equation}
In fig. 4 we show the numerical solution of the equations of Stokesian dynamics eq. (2.5) with forces given by
  \begin{equation}
\label{5.9}\du{F}(t)=\du{E}_0(t)+\du{H}\cdot(\du{R}(t)-\du{S}_0)
\end{equation}
for $d=5a$, stiffness $\sigma=1$, and amplitude factor $\varepsilon=2$ for the first ten periods. We compare the orbit with the ellipse given by eq. (5.2). The mean swimming velocity and mean power, calculated as time-averages over the last period for values of the amplitude factor in the range $0<\varepsilon<2$, are shown in figs. 1 and 2. The corresponding efficiency is shown in fig. 3. The dashed curves in figures $1-3$ replace figs. 3, 4, and 5 of ref. 1, which were calculated from inappropriate actuating forces. The efficiency is approximately twice as large as calculated in ref. 1.

It is true that in fig. 3 the efficiency for given $\varepsilon$ calculated by the kinematic method is always larger than that calculated by the dynamic method from the limit cycle with actuating forces. However, we must compare the mean swimming velocity for two different strokes of the same mean power. In fig. 5 we plot the power as a function of $\varepsilon$ in the range $1.9<\varepsilon<2$ as calculated by the two different methods. The value $\overline{\mathcal{D}}=52\;\eta\omega^2a^3$ of the mean power  occurs at $\varepsilon_k=1.949$ in the kinematic method, and at $\varepsilon_d=1.970$ in the dynamic method. For these values the mean swimming velocity is found to be $\overline{U}_{sw}=0.0546\;\omega a$ for the elliptical orbit of the kinematic method, and $\overline{U}_{sw}=0.0538\;\omega a$ for the limit cycle of the dynamical method. Thus in the present case the elliptical orbit is the most efficient of the two. This does not exclude that for the same power an orbit with yet higher speed can be found.

At $\varepsilon=1.38$ and for $d=5a$ we have $\overline{U}_{sw}\approx 0.026\;\omega a$ from eq. (3.4) and $\overline{\mathcal{D}}\approx 25.8\;\eta\omega^2 a^3$ from eq. (3.22) for the orbit given by eq. (5.3). This can be compared with the numerical calculation of Alouges et al. \cite{2},\cite{3} on the basis of a Stokes solver. The authors used radius $a=0.05$ mm, and period $T=1$ s. For viscosity of water $\eta=0.01$ poise our calculation yields $\Delta=\overline{U}_{sw}T\approx$ 0.0081 mm and $\overline{\mathcal{D}}T\approx 0.127\times 10^{-12}J$. The latter value is somewhat less than the one given in table 1 of ref. 3, and the displacement agrees well with the value $0.01$ mm of Alouges et al..

 Finally we consider the efficiency calculated from eqs. (3.4) and (3.22) for displacement in relative space of the form eq. (5.2), but with the eigenvector $\vc{\xi}_+$ replaced by $\vc{\xi}=(1,A\exp(i\delta))$ with absolute value $A$ and phase $\delta$. The values of $A$ and $\delta$ can be related to the Stokes parameters of the elliptical orbit \cite{12}. In fig. 6 we show the efficiency for amplitude factor $\varepsilon=2$ and ratio $d/a=5$ as a function of $A$ and $\delta$. The maximum is not very pronounced.

\section{\label{6}Discussion}

The swimming performance of an assembly of spheres as a function of the amplitude of a chosen stroke can be studied in a purely kinematic formulation. From eq. (3.4) we find the instantaneous swimming velocity, and from eq. (3.22) we find the instantaneous rate of dissipation or power. The mean swimming velocity and the mean power follow by averaging over a period. The ratio of these two quantities yields the efficiency of the stroke.

Alternatively one may use a dynamic approach \cite{1},\cite{11} in which the swimmer is modeled as a set of spheres bound harmonically to equilibrium positions and with harmonic interactions. The spheres are subject to actuating forces which sum to zero. The corresponding swimming motion may be found as the limit cycle of the solution of the equations of Stokesian dynamics. The mean swimming velocity and the mean power may be found numerically from the limit cycle.

We have shown in sect. 5 that for a collinear three-sphere swimmer the two methods lead to similar results over a wide range of amplitude, provided that for small amplitude the actuating forces correspond to the chosen kinematic stroke. We have chosen the latter to be the optimal one at small amplitude, as determined from the velocity matrix and the power matrix of the bilinear theory.

The kinematic method is the more straightforward one, since it does not require numerical solution of the equations of Stokesian dynamics. The dynamic approach has the advantage that it provides a physical model of the swimmer. It will be of interest to explore the difference in efficiency for given stroke or given actuating forces as a function of amplitude factor for more sophisticated model swimmers, with actuating forces chosen to agree with the optimal stroke at small amplitude.

\newpage

\section*{Figure captions}

\subsection*{Fig. 1}
Plot of the reduced mean swimming velocity $\overline{U}_{sw}/(\varepsilon^2\omega a)$ for $d=5a$ as a function of the amplitude $\varepsilon$ as calculated by the kinematic method (solid curve), and by the dynamic method with stiffness parameter $\sigma=1$ (dashed curve).

\subsection*{Fig. 2}
Plot of the reduced mean swimming power $\overline{\mathcal{D}}/(\varepsilon^2\eta\omega^2 a^3)$ for $d=5a$ as a function of the amplitude $\varepsilon$ as calculated by the kinematic method (solid curve), and by the dynamic method with stiffness parameter $\sigma=1$ (dashed curve).

\subsection*{Fig. 3}
Plot of the efficiency $E_T=\eta\omega a^2\overline{U}_{sw}/\overline{\mathcal{D}}$ for $d=5a$ as a function of the amplitude $\varepsilon$ as calculated by the kinematic method (solid curve), and by the dynamic method with stiffness parameter $\sigma=1$ (dashed curve).

\subsection*{Fig. 4}
Plot of the orbit in the $r_1r_2$ plane calculated from the equations of Stokesian dynamics for $d=5a,\;\varepsilon=2,\;\sigma=1$ for ten periods. The initial values correspond to Eq. (5.8) and the forces follow from eq. (5.9). We also plot the elliptical orbit for $d=5a,\;\varepsilon=2$ (dashed curve).

\subsection*{Fig. 5}
Plot of the mean swimming power $\overline{\mathcal{D}}/(\eta\omega^2 a^3)$ for $d=5a$ as a function of the amplitude $\varepsilon$ in the range $1.9<\varepsilon<2$ as calculated by the kinematic method (solid curve), and by the dynamic method with stiffness parameter $\sigma=1$ (dashed curve).

\subsection*{Fig. 6}
Plot of the efficiency $E_T=\eta\omega a^2\overline{U}_{sw}/\overline{\mathcal{D}}$ calculated by the kinematic method for the elliptical orbit in the $r_1r_2$ plane given by eq. (5.2) for $d=5a$ with $\varepsilon=2$ and $\vc{\xi}_+$ replaced by $\vc{\xi}=(1,A\exp(i\delta))$ as a function of amplitude $A$ and phase $\delta$.

\newpage
\setlength{\unitlength}{1cm}
\begin{figure}
 \includegraphics{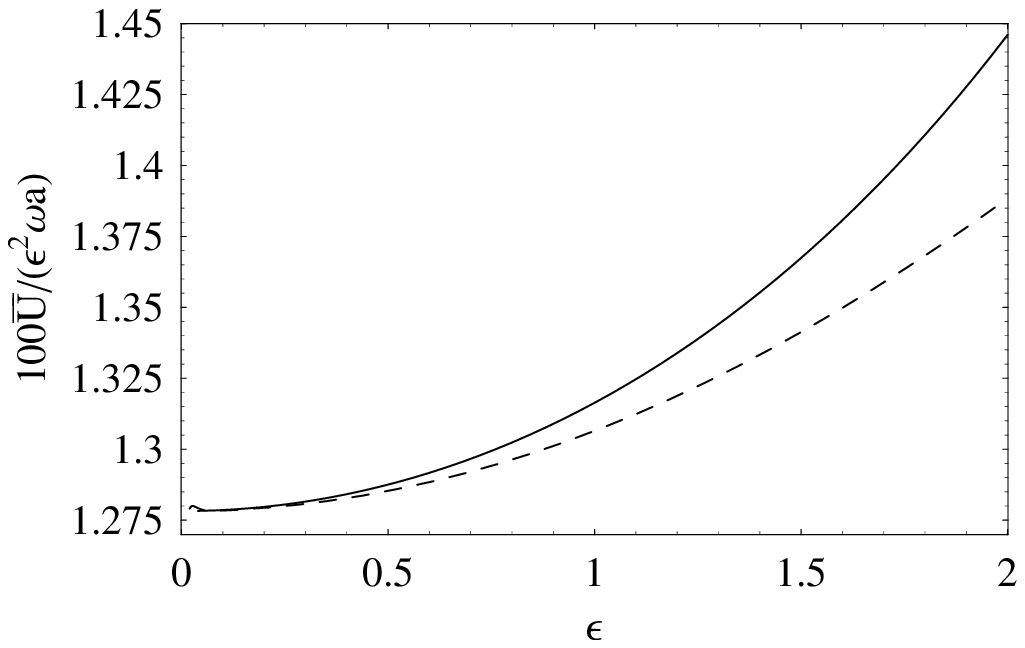}
   \put(-9.1,3.1){}
\put(-1.2,-.2){}
  \caption{}
\end{figure}
\newpage
\clearpage
\newpage
\setlength{\unitlength}{1cm}
\begin{figure}
 \includegraphics{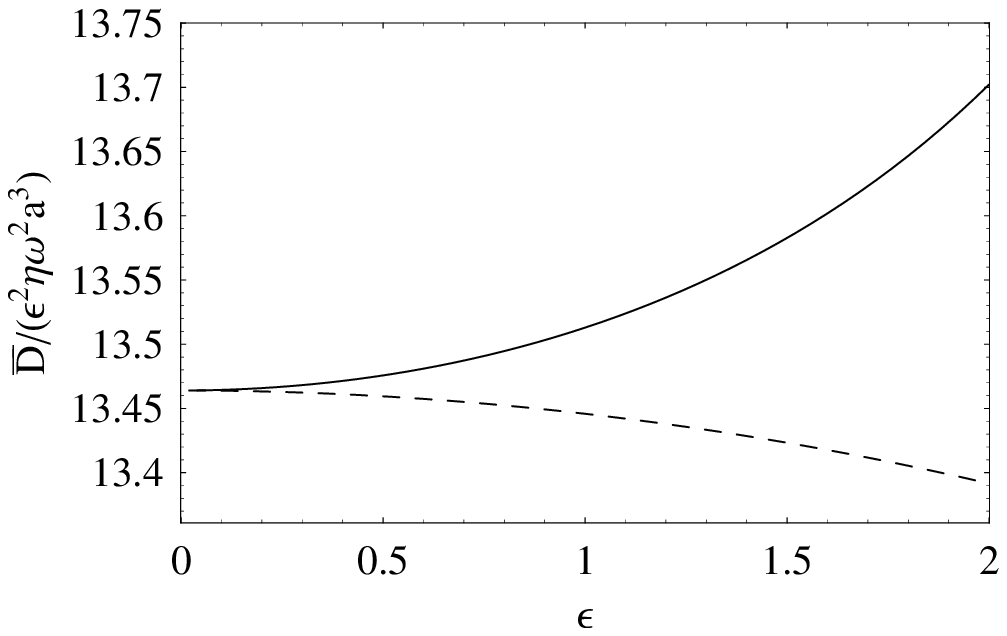}
   \put(-9.1,3.1){}
\put(-1.2,-.2){}
  \caption{}
\end{figure}
\newpage
\clearpage
\newpage
\setlength{\unitlength}{1cm}
\begin{figure}
 \includegraphics{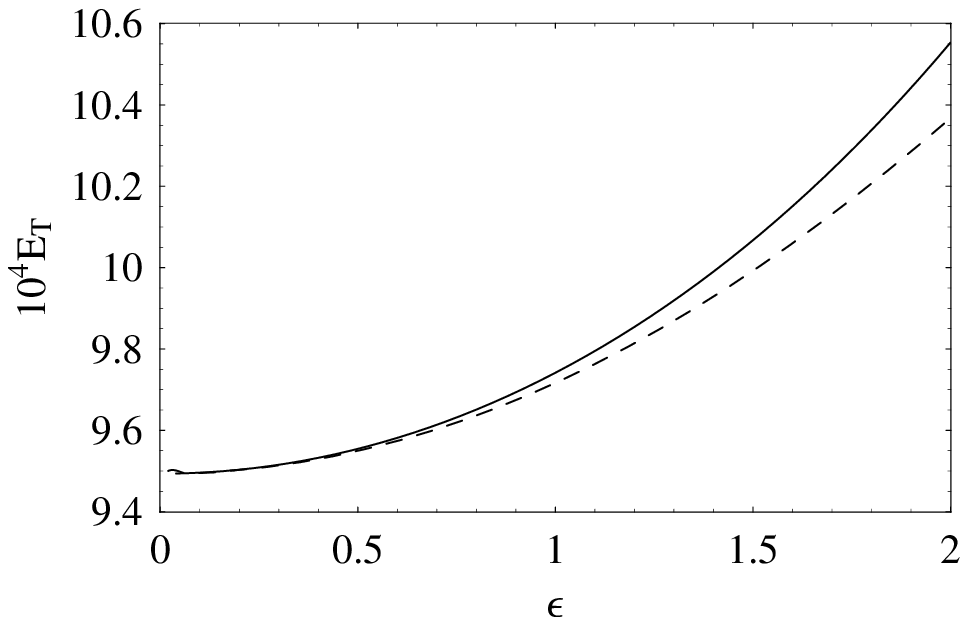}
   \put(-9.1,3.1){}
\put(-1.2,-.2){}
  \caption{}
\end{figure}
\newpage
\clearpage
\newpage
\setlength{\unitlength}{1cm}
\begin{figure}
 \includegraphics{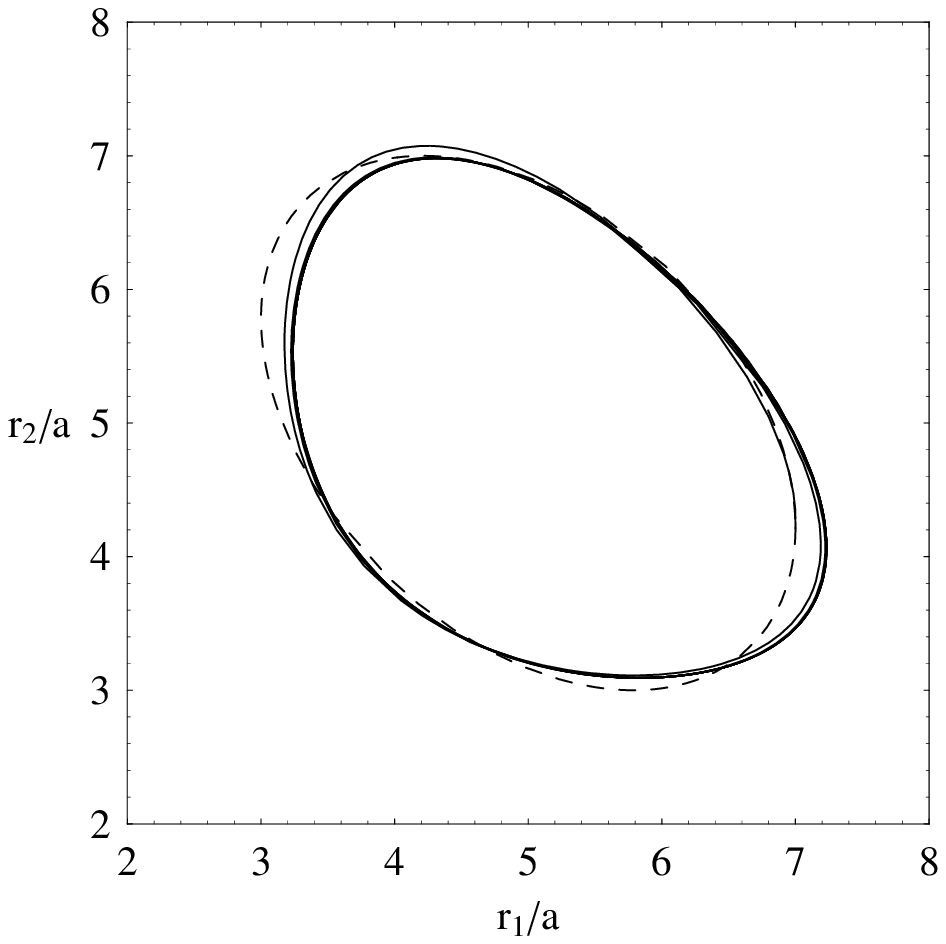}
   \put(-9.1,3.1){}
\put(-1.2,-.2){}
  \caption{}
\end{figure}
\newpage
\clearpage
\newpage
\setlength{\unitlength}{1cm}
\begin{figure}
 \includegraphics{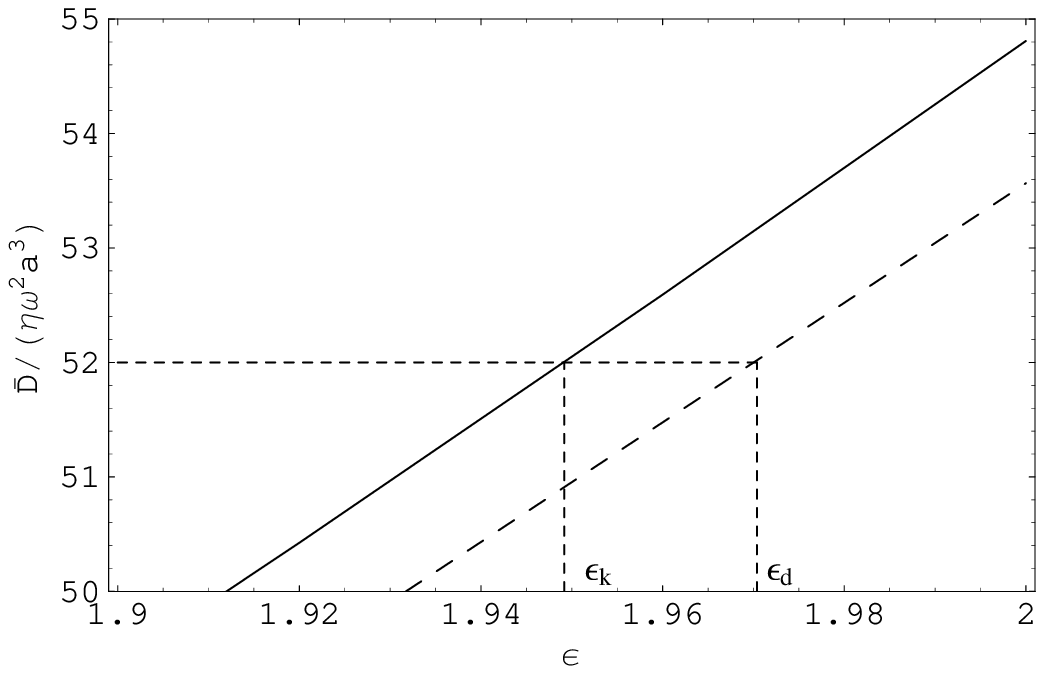}
   \put(-9.1,3.1){}
\put(-1.2,-.2){}
  \caption{}
\end{figure}
\newpage
\clearpage
\newpage
\setlength{\unitlength}{1cm}
\begin{figure}
 \includegraphics{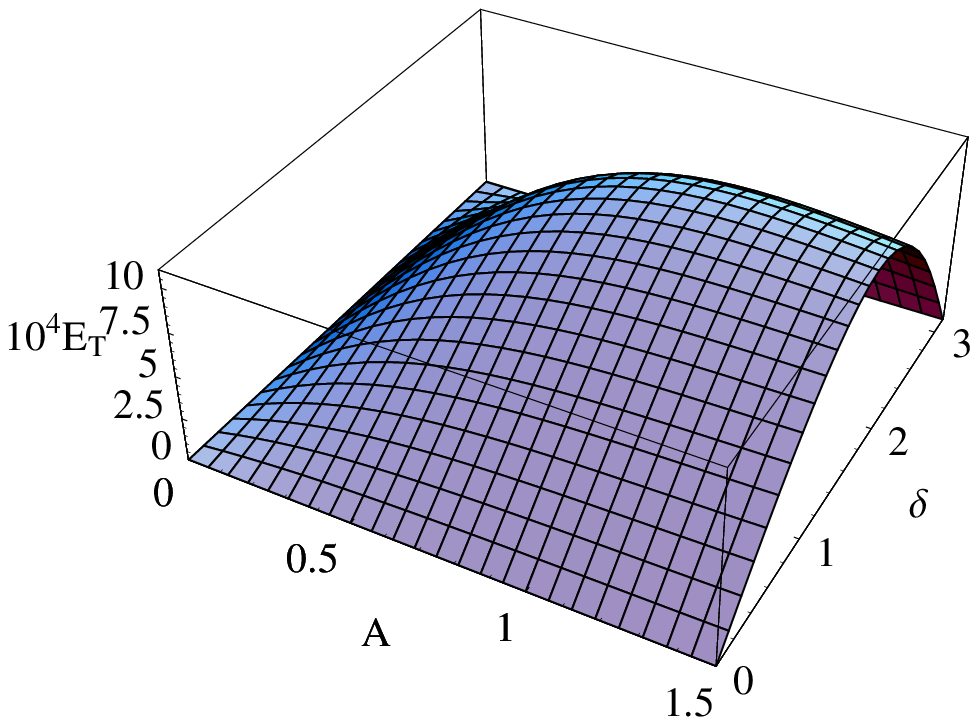}
   \put(-9.1,3.1){}
\put(-1.2,-.2){}
  \caption{}
\end{figure}
\newpage


\newpage

\begin{thebibliography}{99}

\bibitem{1}B. U. Felderhof, Eur. Phys. J. E \vol{37},110 (2014).

\bibitem{2}
F. Alouges, A. DeSimone, and A. Lefebvre, J. Nonlinear Sci. \vol{18}, 277 (2008).

\bibitem{3}
F. Alouges, A. DeSimone, and A. Lefebvre, Eur. Phys. J. E \vol{28}, 279 (2009).

\bibitem{4}
V. A. Vladimirov, J. Fluid Mech. \vol{716}, R1-1 (2013).

\bibitem{5}
A. Najafi and R. Golestanian, Phys. Rev. E \vol{69}, 062901 (2004).

\bibitem{6}
R. Golestanian and A. Ajdari, Phys. Rev. E \vol{77}, 036308 (2008).

\bibitem{7}
J. Happel and H. Brenner, {\it Low Reynolds number hydrodynamics} (Noordhoff, Leyden, 1973).

\bibitem{8}
B. Cichocki, B. U. Felderhof, K. Hinsen, E. Wajnryb, and J. Blawzdziewicz, J. Chem. Phys. \vol{100}, 3780 (1994).

\bibitem{9}
B. Cichocki, M. L. Ekiel-Je\.zewska, and E. Wajnryb, J. Chem. Phys. \vol{111}, 3265 (1999).

\bibitem{10}
M. L. Ekiel-Je\.zewska and E. Wajnryb, in {\it Theoretical Methods for Micro Scale Viscous Flows}, edited by F. Feuillebois and A. Sellier (Transworld Research Network, Kerala, 2009).

\bibitem{11}
B. U. Felderhof, Phys. Fluids \vol{18}, 063101 (2006).

\bibitem{12}
C. F. Bohren and D. R. Huffman, {\it Absorption and Scattering of Light by Small Particles} (Wiley, New York, 1983).



\end{thebibliography}
\end{document}